\documentclass[aps,prd,epsf,showpacs,amsmath,amssymb,11pt]{revtex4}
\usepackage{dcolumn}
\usepackage{bm}

\newcommand{\e}{equation$\;$}

\newcommand{\be}{\begin{equation}}
\newcommand{\ee}{\end{equation}}
\newcommand{\ba}{\begin{eqnarray}}
\newcommand{\ea}{\end{eqnarray}}
\newcommand{\ban}{\begin{eqnarray*}}
\newcommand{\ean}{\end{eqnarray*}}      

\newcommand{\n}[1]{\label{#1}}

\newcommand{\eq}[1]{(\ref{#1})}

\newcommand{\ntr}{\ensuremath{{\nu(t,r)}}}
\newcommand{\str}{\ensuremath{\psi(t,r)}}
\newcommand{\ph}{\ensuremath{{\phi}}}
\newcommand{\dw}{\ensuremath{{d\Omega^2}}}

\newcommand{\E}{{\mathbb E}}

\begin{document}

\title{Static spherically symmetric scalar field spacetimes 
with $C^0$ matching}

\author{Swastik Bhattacharya and Pankaj S. Joshi}
\affiliation{Tata Institute for Fundamental Research, Colaba,
Mumbai 400005, India}

\begin{abstract} 
All the classes of 
static massless scalar field models available 
currently in the Einstein theory of 
gravity necessarily contain a strong curvature naked 
singularity. We obtain here a family of solutions for 
static massless scalar fields coupled to gravity, which 
does not have any strong curvature singularity. This class 
of models contain a thin shell of singular matter, which 
has a physical interpretation. The central curvature singularity 
is, however, avoided which is common to all static 
massless scalar field spacetimes models known so far. 
Our result thus points out that the full class of 
solutions in this case may contain non-singular models, 
which is an intriguing possibility.

\end{abstract}

\pacs{04.20.Jb, 04.20.Dw, 04.70.-s}
\maketitle
\section*{}

Spherically symmetric solutions of Einstein equations for
static massless scalar field configurations have been investigated in 
considerable detail in past. Bergmann and Leipnik 
\cite{Bergman} 
were among the first to construct spherically symmetric static 
solutions for a massless scalar field. They had, however, only
a limited success due to an inappropriate choice of coordinates. 
Around the same time Buchdal 
\cite{Buchdal} 
developed techniques to generate solutions for this system, and also  
Yilmaz 
\cite{Yilmaz} 
and Szekeres 
\cite{Szekeres} 
found some classes of solutions for the static massless scalar 
field configurations in general relativity.

Subsequently, Wyman 
\cite{Wyman} 
systematically discussed these solutions in comoving
coordinates where the energy momentum tensor is in 
diagonal form, and showed 
a general method to obtain solutions in the case when the 
scalar field was allowed to have no time dependence. This gave a 
unified method to obtain most of the solutions obtained earlier.
Also, Xanthopoulos and Zannias 
\cite{Xan} 
gave a class of solutions for time independent scalar fields 
in arbitrary dimensions where the spacetime metric was static.
Further, the static scalar fields conformally coupled 
to gravity have been a subject of immense interest to many 
researchers 
\cite{conformal}, \cite{conformalP}, \cite{conformalB}, \cite{conformalX}.
Static massless scalar fields have also been investigated
in settings more general as compared to spherical symmetry
(see e.g.
\cite{scal}, \cite{scalJ}).

The main interest in these models has been mainly due to 
several interesting properties that these solutions exhibit, as was
pointed out, for example, by the JNW solution
\cite{JNW}.
Mainly these properties refer to the nature of the 
spacetime singularity and the event horizons in these spacetimes.
There are no trapped surfaces in the model and the  
singularity which is visible at $r=2m$ has interesting 
properties
\cite{JNWsing}.

These features were generalized in
\cite{Chase},
leading to a result that for static massless scalar fields,
the event horizon is always singular in asymptotically flat
spacetimes, and that these results are not necessarily restricted
to spherically symmetric models only. In this sense, a study
of static massless scalar fields coupled to gravity
provides some important insights into the global structure
of these spacetimes, and also it gives us  
useful information on the nature of singularities 
and trapped surfaces.

The point here is that, the vacuum spherically symmetric
model is the Schwarzschild solution, which is a black hole
with an event horizon covering the singularity. However, 
an introduction of a smallest
scalar field in the model radically changes the 
causal properties of the model, making the horizon and trapped
surfaces to disappear and the spacetime singularity 
is visible. It is thus a matter of interest to examine if
this class of models admit any singularity free solutions, in order
to decide if the presence of a non-vanishing scalar field always 
causes a naked singularity. While this issue is examined here,
in the process we also find a new class of models for
static massless scalar field, indicating the possibility 
of non-singular massless static scalar field models.

One of the main features of the Wyman class of solutions,
which is the currently available most general class of models 
in the case under consideration, is that there exists in 
these spacetimes a central singularity which is naked. 
In the present note, we report a class of solutions where there 
is no such strong curvature singularity. However, 
a $C^0$ matching is necessary to achieve this, and as a 
result there is a shell which has singular matter. Many 
examples of this type of matching of spacetimes are available 
in the literature. Usually, this type of singularity can 
be given a physical interpretation, unlike the strong 
curvature singularity, and hence it is not considered to be 
pathological as the former. This type of 
matching conditions were first introduced by Lanczos and Israel 
\cite{Israel}. Later on this has been used by many other 
authors (for a nice review on the topic, see e.g. 
\cite{Vickers}). 
The thin shell formalism has also been used for 
static spacetimes 
\cite{Date}. 
In our case also, a $C^0$ matching is performed 
in a static case. This class of solutions presented here 
is different from the Wyman class of solutions and 
does not have a naked curvature singularity.

We consider here a four-dimensional spacetime 
manifold which has spherical symmetry.
The massless scalar field $\ph(x^a)$ on such a spacetime manifold 
$(M, g_{ab})$ is described by the Lagrangian, 
${\cal L}=-\frac{1}{2}\ph_{;a}\ph_{;b}g^{ab}.$
The corresponding Euler-Lagrange equation is then given by, 
$\ph_{;ab}g^{ab}=0$,
and the energy-momentum tensor for the scalar field, as calculated 
from this Lagrangian, is given as 
\be
T_{ab}=\ph_{;a}\ph_{;b}-\frac{1}{2}g_{ab}\left(\ph_{;c}\ph_{;d}
g^{cd}\right).
\n{emt}
\ee
Let us consider the massless scalar field which is a 
{\it Type I} matter field
\cite{haw},  
{\it i.e.}, the energy-momentum tensor admits one timelike 
and three spacelike eigen vectors. At each point $q\in M$, we can 
express the tensor $T^{ab}$ in terms of an orthonormal basis 
$(\E_0,\E_1,\E_2,\E_3)$, 
where $\E_0$ is a timelike eigenvector with the eigenvalue $\rho$
and $\E_{\alpha}$ $(\alpha=1,2,3)$ are three spacelike eigenvectors 
with eigenvalues $p_\alpha$. The eigenvalue $\rho$ represents 
the energy density of the scalar field as measured by an observer 
whose world line at $q$ has an unit tangent vector $\E_0$, and 
the eigenvalues $p_\alpha$ represent the principal pressures 
in three spacelike directions $\E_\alpha$.

We choose the spherically symmetric coordinates 
$(t,r,\theta,\phi)$ along the eigenvectors $(\E_0,\E_\alpha)$, 
such that the reference frame is {\it comoving}, as was 
chosen by 
\cite{Wyman} and also
\cite{Xan}. 
The general spherically symmetric metric can now be 
written as,
\begin{equation}
ds^2= e^{2\ntr}dt^2-e^{2\str}dr^2-R^2(t,r)\dw,
\label{metric}
\end{equation}
where $\dw$ is the metric on a unit 2-sphere and 
we have used the two gauge freedoms of two variables, namely, 
$t'=f(t,r)$ and $r'=g(t,r)$, to make the $g_{tr}$ term in the 
metric and the $T_{tr}$ component of the energy-momentum tensor
 of the matter field to vanish.
Thus the energy-momentum tensor has a diagonal form.
We note that we still have two scaling freedoms of one variable
available, namely $t \to f(t)$ and $r \to g(r)$. We note
that the variable $R$ represents the physical radius.

We have for spherical symmetry $\ph=\ph(t,r)$,
and from  \e\eq{emt} we see that $T_{10}=\phi' \dot{\phi}=0$. 
So we have necessarily $\ph(t,r)=\ph(t)$ or $\phi(t,r)=\phi(r)$, 
with the energy-momentum tensor being diagonal. 
For the metric (\ref{metric}),
and using the following definitions,
\begin{equation}
G(t,r)=e^{-2\psi}(R^{\prime})^{2},\;\; H(t,r)=e^{-2\nu} (\dot{R})^{2}\;,
\n{eq:ein5}
\end{equation}
\begin{equation}
F=R(1-G+H)\;,
\label{eq:ein4}
\end{equation}
we can write the independent Einstein equations for the 
spherically symmetric massless scalar field (in the units $8\pi G=c=1$) 
as below (see \cite{GJ1}, \cite{GJ2}),
\begin{equation}
 \rho=\frac{F'}{R^2 R'},
\end{equation}
\begin{equation}
 P_r=-\frac{\dot{F}}{R^2 \dot{R}},
\end{equation}
\begin{equation}
 \nu'(\rho+P_r)=2(P_\theta-P_r)\frac{R'}{R}-P_r'\label{e3},
\end{equation}
\begin{equation}
 -2\dot{R}'+R'\frac{\dot{G}}{G}+\dot{R} \frac{H'}{H}=0\label{e1}, 
\end{equation}

In the above, the function $F(t,r)$, 
also called the Misner-Sharp mass, has the 
interpretation of the mass for the matter field, in that it represents 
the total mass contained
within the sphere of coordinate radius $r$ at any given time $t$. 
As noted above, in the static case, the metric components 
$g_{\mu\nu}$s are functions of $r$ only necessarily, but  
the scalar field $\phi$ itself can be in general either 
$r$ or $t$ dependent.  In the case when $\phi=\phi(r)$,
which we consider here, 
the components of the energy-momentum tensor are given by,
\begin{equation}
T^t_t=-T^r_r=T^{\theta}_{\theta}=T^{\phi}_{\phi}=\frac{1}{2}e^{-2\psi}\ph'^2
\end{equation}
It follows that the equation of state in this case, 
which relates the scalar field energy density and pressures 
is thus given by $\rho=P_r=-P_{\theta}$.
As noted by Wyman \cite{Wyman}, there can be 
a class of static spacetimes where $\phi=\phi(t)$ 
also. But we would consider here only the class 
for which $\phi=\phi(r)$, which describes many earlier 
known interesting 
solutions for static scalar field spacetimes.

We shall now consider the static spacetimes, when $\phi=\phi(r)$ 
and $g_{\mu\nu}= g_{\mu\nu}(r)$. 
The Einstein equations given above then reduce 
to the following set of equations,
\begin{equation}
\frac{1}{2}e^{-2 \psi} \phi'^2 = \frac{F'}{R^2 R'},
\end{equation}
\begin{equation}
\frac{1}{2}e^{-2\psi} \phi'^2 = e^{-2\psi}(\frac{R'^2}{R^2}+\frac{2R'\nu'}{R})
-\frac{1}{R^2},
\end{equation}
\begin{equation}
\phi'' = (\psi'-\frac{2R'}{R}-\nu')\phi'\label{s1}.
\end{equation}
\begin{equation}
e^{-2\psi} R'^2 = 1-\frac{F}{R} 
\end{equation}
In the above, the equation \eqref{s1} can be integrated once 
with respect to $r$ to give
\begin{equation}
\phi'=\frac{e^{\psi-\nu+a}}{R^2}\label{fld},
\end{equation}
where $a=const.$
Eliminating now $\phi'$ from these equations gives,
\begin{equation}
\frac{1}{2}\frac{e^{-2\nu+2a}}{R^2} R' = F'\label{s2},
\end{equation}
\begin{equation}
\frac{1}{2}\frac{e^{-2\nu+2a}}{R^4} = e^{-2\psi}(\frac{R'^2}{R^2}+\frac{2R'\nu'}
{R})-\frac{1}{R^2} \label{s3}
\end{equation}
\begin{equation}
e^{-2\psi} R'^2 = 1-\frac{F}{R} \label{s4}
\end{equation}
We note that there is still a freedom left to transform the 
radial coordinate $r$, and thus the number of unknown variables is 
reduced to three in the three equations above. This freedom is just a 
coordinate transformation of the form 
\begin{equation}
 r\to \chi(r),
\end{equation}
which is allowed by the spherical symmetry of the spacetime.

Apart from the above equations, we can also obtain a 
useful first integral, which is actually contained 
in the previous equations and it is advantageous to use it. 
The Einstein equations in this case can be written 
in the form $R_{\mu\nu} = \phi,_\mu \phi,_\nu$.
This implies $R_{00}=0$, from which we get
\begin{equation}\label{1st}
\nu' = \frac{h e^{\psi-\nu}}{R^2}
\end{equation}
where the quantity $h$ is a constant. From\eqref{1st} and 
\eqref{fld}, we get, 
\begin{equation}
\nu=\frac{\alpha\phi}{2}+C_1\label{2nd}
\end{equation}
where $\alpha= he^{-a}$ and $C_1 $ are constants.

To examine the Einstein equations 
above, we now define a function $f(R)$ as below,
\begin{equation}
f(R),_R = \frac{e^{-2\nu}}{2R^2} \label{gic}
\end{equation}
The above is a general definition, and not any
assumption, because the metric functions here depend
on $r$ only.
Using this in \eqref{s2}, we get,
\begin{equation}
F=e^{2a}f(R)+C_2 
\end{equation}
We can choose $C_2=0$, which gives $F=e^{2a}f(R)$. 
Using this in \eqref{s4}, we get
\begin{equation}
e^{-2\psi} R'^2 = 1-\frac{e^{2a}f(R)}{R}
\end{equation}
Also, from \eqref{gic} we get,
\begin{equation}
-2\nu'=(\frac{f,_{RR}}{f,_R}+\frac{2}{R})R'
\end{equation}

Taking $e^{2a}=1$ and $C_2=0$ for simplicity and clarity 
of presentation, the last two equations, together 
with \eqref{s3} give
\begin{equation}
 R(f,_R)^2 = (f-R)(f,_R+Rf,_{RR})-Rf,_R \label{gi}
\end{equation}
The equation above holds true in generality and 
we have not made here any assumption of a special 
coordinate condition or no specific radial coordinate 
choice has been made. 
The above represents clearly the main Einstein 
equation in the static case when $\phi=\phi(r)$.
The solutions to the same give the classes of 
allowed static massless scalar fields models 
in general relativity.  The above is a non-linear
ordinary differential equation of second order which is 
in general difficult to solve fully.

We now consider a particular solution of 
\eqref{gi}, which is given by 
$f(R)=-\frac{1}{R}$, which solves the above as 
is easy to check by inspection. 
This solution of $f(R)$ gives a class of solutions 
to the Einstein equations as will be shown now.
It should be noted here that for $f(R)=-\frac{1}{R}$, 
we have from \eqref{gic}, $e^{-2\nu}=2$ and 
so $\nu= const$. Then from \eqref{1st}, we get $h=0$, 
and therefore $\alpha=0$ in \eqref{2nd}. 
This shows that $\phi$ can still be a non-constant 
function of $r$ even when $\nu$ is constant.

From now on, we focus here only
the class given by the condition $f(R) = -1/R$. 
To write down the solution explicitly and to
specify the same in terms of the metric components, 
a choice of the radial coordinate $r$ has to 
be made now. 
We note here that in the above consideration, 
there exists 
a freedom of choosing $\phi(r)$, and therefore a 
choice for the same corresponds to actually making a choice 
of the radial coordinate $r$. Only after we have made a choice of 
$\phi(r)$, would that scaling be fixed.

In general, for the case $f= -\frac{1}{R}$, from earlier
equations we have,
\begin{equation}
e^\psi = \frac{1}{\sqrt{2}}R^2 \phi'(r) \label{psi}
\end{equation}
Then, the Einstein equation $e^{-2\psi} R'^2 = 1-\frac{F}{R}$ 
implies that we have, 
$$\frac{2R'^2}{\phi'^2(r)}=R^2(R^2+1).$$ 
By solving the above equation we get,
\begin{equation}
 R= \frac{1}{\sinh{(\pm \frac{1}{\sqrt{2}}\phi(r))}} \label{req1}
\end{equation}

It thus follows that we can now write down explicitly 
all the metric components, in terms of the function $\phi(r)$ 
and $\phi'(r)$ above, thus giving a full solution. 
The metric coefficient $\nu$ can be found out from \eqref{gic}.
The metric component $R^2$ can be found out from 
\eqref{req1}. Finally the function $\psi$ 
is also known from \eqref{psi}. This completes the solution.

The function $\phi(r)$ here can be viewed as a free 
function that we are actually allowed choose, and the 
energy conditions are always respected for all such choices.
As an example, putting $\phi(r)=\frac{1}{r}$, we 
can recover a particular solution from the Wyman class
\cite{Wyman},  
which is given by, 
\begin{equation}
 ds^2= \frac{1}{2}dt'^2- [\frac{r^{-1}}{\sinh{(\frac{1}
{\sqrt{2}} r^{-1})}}]^4 dr^2- 
[\frac{r^{-1}}{\sinh{( \frac{1}{\sqrt{2}}
r^{-1})}}]^2 r^2 d\Omega^2 \label{wyman2}
\end{equation}
We note that the range of $\phi$ is from
zero to infinity in this case.

In general, we can in fact recover the entire Wyman 
class by this procedure. Each solution of \eqref{gi} with 
$\phi(r)=\frac{1}{r}$ gives one solution from 
the Wyman class. The solution set of \eqref{gi} 
with this choice of $\phi(r)$ gives the whole Wyman class 
of solutions.
We note here that, in general, the choice 
of the function $\phi(r)$ is not just a gauge choice. 
For example, if we take $\phi(r) = -\frac{1}{r}$,
the range of $\phi$ has an upper bound, but no 
lower bound and it goes from zero to negative infinity 
and the upper bound can be changed from zero to any other 
number also. In the choice that Wyman made, the range 
of $\phi$ is different, and is necessarily restricted from  
below in that it goes from zero to positive infinity. 
Thus, in general the solution would be different 
from the Wyman class of models.


For the metric given by \eqref{wyman2}, the 
Ricci scalar $R_c$ is given by 
\begin{equation}
 R_c= -\frac{2}{R^4} \label{RicciS}
\end{equation}
Therefore there is a curvature singularity at the 
center $R=0$. In what follows, we construct a class of
solutions where the central singularity is absent, 
by choosing a specific form of $\phi(r)$. Before proceeding 
further, however, we need to calculate the 
quantity $P_l= \int e^\psi dr$. This would be necessary 
to find out the proper length 
between two shells on any $t= constant$ hypersurface. 
\begin{equation}
P_l(r)=\int e^\psi dr = \frac{2\sqrt{2}}{1-e^{\sqrt{2}\phi(r)}} \label{properl}
\end{equation}

Our purpose here is to construct and find a solution  
without any strong curvature singularity in the spacetime. 
Towards that purpose, we need to 
remove the central singularity $R=0$, which for example 
exists in all other models, as discussed above.   
One way to achieve this is to construct a solution 
such that the physical radius $R$ in fact does not 
vanish. To do this, we first notice,
from \eqref{req1}, that $R \to 0$ 
when $\phi(r) \to \pm\infty$. Also from \eqref{properl}, 
we see that $P_l$ diverges when 
$\phi \to 0$. Since we are considering a static 
solution without any strong curvature singularity here, 
we can consider any $t= constant$ hypersurface where the 
physical radius $R$ must not vanish. In that case, there
is no strong curvature central singularity in the spacetime.
Further, since the spacetime is static here, it 
will be inextendible provided the proper length from 
any point on the spacelike hyperspace (any $t=const$ surface)
to the outer boundary of the hypersurface is infinite. 
This ensures the regularity of the solution.
If both of these requirements are to be satisfied, 
then $\phi$ cannot diverge and must 
go to zero twice in a range of the radial 
coordinate $r$. This implies that $\phi(r)$ must have an 
extrema somewhere in that range of $r$ where $\phi'(r)=0$.

There are in fact an infinite number of 
functions which satisfy these criteria, as required
above, and so all of them are 
equivalent in this regard. Therefore, we make 
a simple choice here as given by,
\begin{equation}
 \phi(r)= (r-a)(b-r) \label{phichoice}
\end{equation}
For this choice, $\phi$ has a maxima at 
$r= \frac{(a+b)}{2}$ where $\phi'(r)=0$. From \eqref{psi}, 
it follows that $e^\psi=0$ at this point, 
which means that the comoving coordinate system 
breaks down there. However, in the limit of 
$r \to \frac{(a+b)}{2}$, the curvature $R_c$ remains finite. 
Also, the proper length between the shells 
$r= \frac{(a+b)}{2}$ and $r= b$ is infinite. 
So it is seen that this coordinate system covers 
the entire domain,
\begin{equation}
 R_{min}\le R\le \infty \label{domain},
\end{equation}
where 
$R_{min}= \frac{1}{\sinh{[\frac{1}{\sqrt{2}}\frac{(b-a)^2}{4}}]}$.

It is clear that the spacetime can be extended through 
the hypersurface $R= R_{min}$. We do this by joining two 
identical domains $R_{min}\le R\le \infty $ together at the 
hypersurface $R= R_{min}$.
In this case, there is no central curvature singularity 
in the 
spacetime and two such identical domains are glued together
to give the full spacetime. 
To examine the matching at this joint, 
we need to find out the extrinsic curvature at 
the hypersurface $r= \frac{(a+b)}{2}$. 
First, we rescale the time coordinate $t' \to \tau$, 
such that $d\tau^2= \frac{1}{2}dt'^2$.
We consider the orthonormal frame given by 
\begin{eqnarray}
n_\mu&=& (0,\rvert e^{2\psi}\rvert,0,0)\\
e_{(\tau)}^\mu&=&(1,0,0,0), \\
e_{(\theta)}^\mu&=&(0,0,1/R,0), \\
e_{(\varphi)}^\mu&=&(0,0,0,1/R\sin\theta).
\end{eqnarray}
The extrinsic curvature of the hypersurface is then given by 
\begin{equation}
K_{(A)(B)}=-e_{(A)}^\mu e_{(B)}^\nu\nabla_\mu n_\nu 
=n_\nu e_{(A)}^\mu\nabla_\mu e_{(B)}^\nu,
\end{equation}
with $A,B= \tau,\theta, \phi$.

In this case, 
\begin{equation}
K_{(\theta)(\theta)}= \frac{|e^{2\psi}|}{e^{2\psi}}
\frac{\coth (\frac{1}{\sqrt{2}}(r-a)(b-r))}{R^2},
\end{equation}
and 
\begin{equation}
K_{(\varphi)(\varphi)}= \frac{|e^{2\psi}|}{e^{2\psi}}
\frac{\coth (\frac{1}{\sqrt{2}}(r-a)(b-r))}{R^2 },
\end{equation}
and the other components vanish.  Now we note that, 
\begin{equation}
\lim_{r\rightarrow \frac{(a+b)}{2} \pm 0}
\frac{|e^{2\psi}|}{e^{2\psi}}\coth (\frac{1}{\sqrt{2}}(r-a)(b-r))=
\mp \coth\left[\frac{(b-a)^2}{4\sqrt{2}}\right],
\end{equation}
and thus the extrinsic curvature of the 
hypersurface $r=r_{\rm min}$ is multi-valued. 
This means that there is a distributional source on 
this hypersurface, which we now calculate. 
The stress energy tensor of this source is 
given by $T_{\mu \nu}= S_{\mu \nu}\delta(\eta)$, where $\eta$ 
is the Gaussian normal coordinate. 
\begin{equation}
 S_{(A)(B)}= e^\mu_{(A)}e^\nu_{(B)}= 
[(K^+_{(A)(B)}-h_{(A)(B)}trK^+)-(K^-_{(A)(B)}-h_{(A)(B)}trK^-) \label{surfaces}
\end{equation}
where $\lim_{r\rightarrow \frac{(a+b)}{2} \pm 0}K_{(A) (B)}= K^{\pm}_{(A) (B)}$ and 
$h_{(A) (B)}$ is the induced metric on the hypersurface $R=R_{min}$.\\
From this it follows that, 
\begin{equation}
 S_{(\tau)(\tau)}= -2 \coth{\frac{(b-a)^2}{4\sqrt{2}}}
= -S_{(\theta)(\theta)}= -S_{(\phi)(\phi)}
\end{equation}
and the other components are zero.

We note here that the surface energy density of 
this hypersurface is negative. We can, however, see now 
that the strong curvature singularity is removed 
in this case by introducing a $C^0$ matching at this 
hypersurface. The solution considered here is topologically 
different from the Wyman class of solutions, as there is 
no central shell in this case. There is a thin shell of singular 
matter in this case with a negative energy density. However,
if we consider a sufficiently thick shell which includes the 
singular thin shell, then the mass inside that thick shell 
is positive. The negative energy thin shell then reduces 
the total mass content of the thick shell. 
In fact, this allows one to give a physical interpretation
to the $C^0$ matching. While there is a mild singularity 
as discussed above at the joining surface, it can be 
given a sound physical interpretation (unlike the strong 
curvature singularity), as already discussed and 
used in the literature
(see e.g. \cite{poisson}).

We also note here the fact that the spacetime we 
construct here is asymptotically flat. This can be seen 
in the following way. We consider only one part of 
the two identical domains, corresponding to 
$R_{min}<R\le \infty$. Restricting ourselves only to this 
one part of the spacetime, we can consider the 
coordinate transformation so that the scalar field 
$\phi(\bar{r})= \frac{1}{\bar{r}}$. This is the gauge 
that Wyman used to write down a class of solutions. 
However, our solution is different from that solution, 
because the coordinate system used by Wyman 
covers only a part of the spacetime manifold of the
solution given by us. Also, our solution does not 
have any strong curvature central singularity, unlike 
the Wyman model. In the Wyman coordinate system, the 
metric of our solution takes the form given in 
\eqref{wyman2}. From this expression, it is seen that  
in the limit $\bar{r} \to \infty$, $R \to \infty$, and 
then the metric becomes Minkowskian. So the 
part of the spacetime corresponding to $R_{min}<R<\infty$, 
is asymptotically flat. Similarly, the other part of 
the spacetime can also be shown to be asymptotically 
flat.

\begin{acknowledgements}
 We would like to thank Ken-ichi Nakao for his helpful 
suggestions.
\end{acknowledgements}

\end{document}